\documentclass[prl,twocolumn]{revtex4}

\usepackage{graphicx}
\usepackage{dcolumn}
\usepackage{bm}
\newcommand{\no}{\noindent}

\newcommand{\bo}{\raise-1mm\hbox{\Large$\Box$}}
\newcommand{\bra}[1]{\langle #1|}
\newcommand{\ket}[1]{|#1\rangle}
\newcommand{\braket}[2]{\langle #1|#2\rangle}
\newcommand{\rr}{\mathbf{r}}

\usepackage{epsfig}
\usepackage{latexsym}
\usepackage{mathrsfs}
\usepackage{setspace}
\usepackage{amsmath}
\usepackage{floatflt}
\usepackage{wrapfig}
\long\def\symbolfootnote[#1]#2{\begingroup
\def\thefootnote{\fnsymbol{footnote}}\footnote[#1]{#2}\endgroup}

\begin{document}

\title{Manifestations of the Roton Mode in Dipolar Bose-Einstein Condensates}

\author{Ryan M. Wilson\email{rmw@quantum.colorado.edu}}
\affiliation{JILA and Department of Physics, University of Colorado, Boulder, Colorado 80309-0440, USA}
 
\author{ Shai Ronen}
\affiliation{JILA and Department of Physics, University of Colorado, Boulder, Colorado 80309-0440, USA}

\author{Han Pu}
\affiliation{Department of Physics and Astronomy, and Rice Quantum Institute, Rice University, Houston, Texas 77251-1892, USA}

\author{ John L. Bohn}
\affiliation{JILA and Department of Physics, University of Colorado, Boulder, Colorado 80309-0440, USA}
\date{\today}

\begin{abstract}
We investigate the structure of trapped Bose-Einstein
condensates (BECs) with long-range anisotropic dipolar interactions.
We find that  a small perturbation in the trapping potential can
lead to dramatic changes in the condensate's density profile for
sufficiently large dipolar interaction strengths and trap aspect
ratios.  By employing perturbation theory, we relate
these oscillations to a previously-identified ``roton-like'' mode in
dipolar BECs.  The same physics is responsible for radial density
oscillations in vortex states of dipolar BECs that have been predicted
previously.
\end{abstract}

\maketitle

The study of ultracold atomic and molecular gases is 
notable for its connection to
%,and possible illumination of%, 
denser condensed matter systems.  Ultracold gases can show a strong resemblance to condensed matter systems such as vortex lattices, superfluids and Mott-insulators.  Part of the
attraction to these analogies is the 
%exquisite% 
ability to control an
ultracold atomic or molecular gas, allowing researchers to explore
regions of parameter space that are difficult to access in a naturally
occurring system.

%A recently elucidated example%
A recent example  of this connection between ultracold
gases and ``conventional'' condensed matter systems arises in dilute
Bose-Einstein condensates (BECs) consisting of dipolar particles.  An
early theoretical study investigated a gas of dipoles that are free to
move in a plane, but are confined in the direction orthogonal to the
plane and that are polarized in this same direction. This system is
predicted to exhibit an anomalous dispersion relation that possesses a
minimum at a characteristic momentum, reminiscent of the roton
dispersion well-known in superfluid $\mathrm{He}$~\cite{Santos03a}.
Moreover, the depth of this minimum is controlled by the strength of
the dipolar interaction (proportional to the square of the dipole moment
and the density).  If this interaction is large enough, the roton
minimum can become degenerate with the ground state.

In experiments, however, the gas is confined, leading to a discrete excitation spectrum rather than a continuous
dispersion relation.  Nevertheless, signatures of the roton excitation
have been identified in calculations with fully three-dimensional trap
geometries~\cite{Ronen07}.  Certain excitations exhibit nodal
structures on the same
%(both radial and angular)% 
%characteristic% 
length scale as the free rotons.  Moreover, the
excitation energies of these modes drop rapidly as the dipolar
interaction strength increases.  
%These results have firmed up the
%formal analogy between roton physics in bulk superfliud $\mathrm{He}$
%and in experimentally accessible dipolar gases. 
We note that the first dipolar BECs (dBECs) have already been created using atomic
$^{52}\mathrm{Cr}$~\cite{Griesmaier05a,KochNature08a,Griesmaier05b,Lahaye08} while molecular BECs (promising far larger dipoles and tunable dipole
moments) are the target of active experimental work.

In this Letter we explore another aspect of roton physics that may be
observable in dipolar gases.  It has previously been suggested that
%``free surfaces'' or %
boundaries in superfliud $^{4}\mathrm{He}$, including vortex cores, should give rise to radial density
oscillations whose length scale is characteristic of the roton
wavelength~\cite{Bohm67,Regge72,Dalfovo92}.  More recently,
calculations of vortex states in a dBEC in a highly oblate trap have exhibited similar radial
structures~\cite{Pu06}, raising the question of the relation between
these structures and rotons in this system (progress has also been made in the understanding of the vortex state in a dBEC in the Thomas-Fermi regime.  However, in this regime the vortex does not manifest a radial ripple~\cite{ODell07,Bijnen07}).

Our objective in the present work is to explore this relationship.
Whereas the complete description of superfluid He is complicated by
%high densities and consequent% 
strong interactions, this is not the
case in a dBEC, where the gas is
 %On the contrary, although the gas must be described by nonlocal interactions, it is nevertheless %
dilute enough
%of low enough density% 
that a mean-field approach should work quite
well~\cite{Bortolotti06}. Indeed, we find that the main effect
generating the radial density oscillations is the perturbation caused by the
centrifugal potential of the vortex state.
%a perturbation of an otherwise smoothcondensate density profile%.  
This perturbation contaminates the ground
state with the lowest-lying excited state, which, in the limit of
strong interactions, is the roton excitation.  We demonstrate this
effect 
%semi-quantitatively% 
by applying a 
%kind of% 
perturbation theory to the nonlinear mean-field equations; the
perturbative approach is in good agreement with our full numerical
calculations.

At very low temperatures, $N$ bosons trapped in an external potential $U(\rr)$ may be described within mean field theory by the nonlocal Gross-Pitaevskii Equation (GPE):
\begin{eqnarray}
\label{GPE}
&& [ -\frac{\hbar^2}{2m}\nabla^2 + U(\rr) + (N-1) \\ \nonumber
&& \times  \int d\rr'\,\Psi^*(\rr') V(\rr-\rr') \Psi(\rr')] \Psi(\rr) = \mu \Psi(\rr),
\end{eqnarray}

\no where $\Psi(\rr,t)$ is the condensate wavefunction (with unit
norm), $\rr$ is the distance
from the trap center, $m$ is the boson's mass, and $V(\rr-\rr')$ is the
two-particle interaction potential.  We consider the case of a
cylindrical harmonic trap, for which the external potential is $U(\rr)
= \frac{1}{2}m\omega_\rho^2(\rho^2+\lambda^2 z^2)$, where
$\lambda=\omega_z/\omega_\rho$ is the trap aspect ratio.  The
interaction potential has the form~\cite{Yi00}
\begin{equation}
\label{V}
V(\rr-\rr') = \frac{4\pi \hbar^2 a_s}{m}\delta(\rr-\rr') + d^2 \frac{1-3\cos^2{\theta}}{|\rr-\rr'|^3}
\end{equation}

\no where $a_s$ is the $s$-wave scattering length, $d$ is the dipole
moment, and $\theta$ is the angle between the vector $\rr-\rr'$ and the dipole axis.  The first term
in $V(\rr-\rr')$ is the familiar contact potential, while the second
term is the long-range anisotropic dipole-dipole potential.  This
potential describes interactions of dipoles that are polarized along
the trap axis, as could be achieved in an experiment by applying a
strong external field.  For the sake of illuminating purely dipolar
effects, we set $a_s=0$ in this work, a limit that can potentially be
achieved experimentally in $^{52}\mathrm{Cr}$~\cite{Werner05}.

Due to the azimuthal symmetry of both the trapping potential and the
dipole-dipole potential, the ground state solutions of Eq.~(\ref{GPE}) may be written in the form $\Psi(\rr,t) = \psi(\rho,z) \,
e^{i k \varphi}$, where $k$ is the quantum number representing the
projection of orbital angular momentum about the trap's
axis~\cite{BEC2003}.  The $k=0$ solutions of Eq.~ (\ref{GPE})
correspond to rotationless BECs, while
the $k=1$ solutions correspond to BECs with singly-quantized
vortices.  The radial structure of the vortex is the same as that of a
rotationless BEC in a trap with a central potential representing the
centrifugal force: indeed, by
%A straightforward way to demonstrate this equivalence is by inserting the form written above into Equation (\ref{GPE}),%
inserting the vortex form written above into Eq.~(\ref{GPE}), one obtains:
%giving the reduced
%GPE for condensates with azimuthal symmetry,
\begin{widetext}\begin{equation}\label{GPE2}
\bigg\{ -\frac{\hbar^2}{2m}\nabla_\rho^2 -
\frac{\hbar^2}{2m}\frac{\partial^2}{\partial z^2} + \frac{\hbar^2
k^2}{2m \rho^2} + \frac{1}{2}m \omega_\rho^2(\rho^2+\lambda^2 z^2) +
(N-1)\int d\rr'\,\psi^*(\rho',z') V(\rr-\rr') \psi(\rho',z') \bigg\}
\psi(\rho,z) = \mu \psi(\rho,z).
\end{equation}\end{widetext}
\no The centrifugal potential $\hbar^2 k^2/2m \rho^2$ is responsible
for the vortex core (i.e., vanishing density at $\rho=0$).

Following the systematic mapping of the structure and stability of
$k=0$ dBECs in oblate traps \cite{Ronen07}, here we undertake to
characterize the structure and stability of a $k=1$ vortex in
dBECs. To characterize the dipolar interaction strength we introduce
the dimensionless parameter $D=(N-1) \frac{m d^2 }{ \hbar^2
a_\mathrm{ho}}$, 
%where $d$ is the dipole moment%,
where $a_\mathrm{ho} = \sqrt{\hbar / m \omega_\rho}$ is the radial
harmonic oscillator length.  
%and $N$ is the number of atoms (or molecules) in the dBEC.%
The dBEC possesses dynamic stability when all of the excited-state
Bogoliubov de Gennes (BdG) eigenenergies are real-valued.  As was
found for $k=0$ condensates in reference~\cite{Ronen07}, we find that
for the vortex state there also exists, at any finite aspect ratio, a
critical value of $D=D_\mathrm{crit}$ above which the $k=1$ dBEC is
dynamically unstable to small perturbations, while below it the vortex
is dynamically stable. We assume that the trap itself is non-rotating,
so that the vortex is not the lowest energy state and therefore is not
thermodynamically stable. However, here we are interested in the
question of the dynamical stability, which is relevant for a closed
system at $T=0$.

\begin{figure}
\includegraphics[width=\columnwidth]{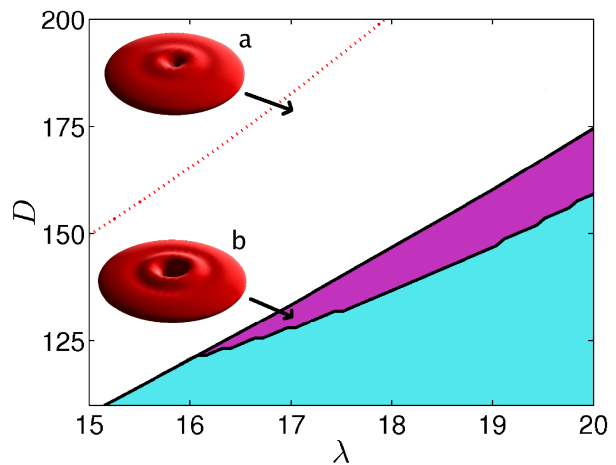}
\caption{\label{fig:stab} The red (thin) dotted line marks the maximum
dipole strength, for a given trap aspect ratio $\lambda$, below which a
rotationless ($k=0$) dBEC is dynamically stable. The colored regions
represent the dynamically stable region for a $k=1$ dBEC, while the
pink (darker) region is where radial oscillations with local minima
are observed.  The inset (a) is an iso-density surface plot of a $k=0$
dBEC perturbed by a small gaussian potential centered on the trap axis, while the inset (b) is an iso-density surface plot of a $k=1$ dBEC.  The presence of radial oscillations is clear in both cases.}
\end{figure}

We find $D_\mathrm{crit}$ for various trap aspect ratios by solving
the linearized BdG equations, as was done in
reference~\cite{Ronen06a}.  Fig.~\ref{fig:stab} illustrates the regions of dynamical stability
for $k=1$ and $k=0$ dBECs
at trap aspect ratios that are relevant to this Letter. We concentrate
here on a specific region in parameter space where we find ripples in
the vortex density profiles, as illustrated in the insets. Whereas previously the ripple was reported for a trap with aspect
ratio $\lambda \sim 100$ and required the existence of a negative
scattering length~\cite{Pu06}, we find that vortices with ripple structure exist
also at milder trap aspect ratios of $\lambda \sim 17$ and with purely dipolar interactions ($a_s=0$).

It is natural to hypothesize that the appearance of the ripple in the
vortex structure is related to a roton mode which is excited by the
centrifugal potential of Eq.~(\ref{GPE2}). This raises the interesting
question, could such a ripple also be observed in the ground (non-vortex)
state of dBEC perturbed by an external potential at the center of the
trap?  Such a perturbation
may be realized experimentally by applying a blue-detuned laser along the trap axis, taking
the form $U'(\rr) = A \exp{(-\rho^2 / 2 \rho_0^2)}$, where $A$ is the
height of the Gaussian and $\rho_0$ is its width.

%As a test case, we consider a dBEC of atomic $^{52}\mathrm{Cr}$.
%$^{52}\mathrm{Cr}$ has magnetic dipole moment $d=6 \mu_\mathrm{B}$,
%the largest of any atom that has been experimentally Bose-condensed.  Additionally,
%$^{52}\mathrm{Cr}$ has an optical dipole transition of 427.6 nm from
%its $^{7}S_3$ ground state to its $^{7}P_3$ excited
%state~\cite{Sugar}.  A blue-detuned laser around this transition could be used to create a repulsive
%potential with no significant absorption and with a minimal spatial extent of about a half-wavelength, ~210 nm. 

\begin{figure}
\includegraphics[width=\columnwidth]{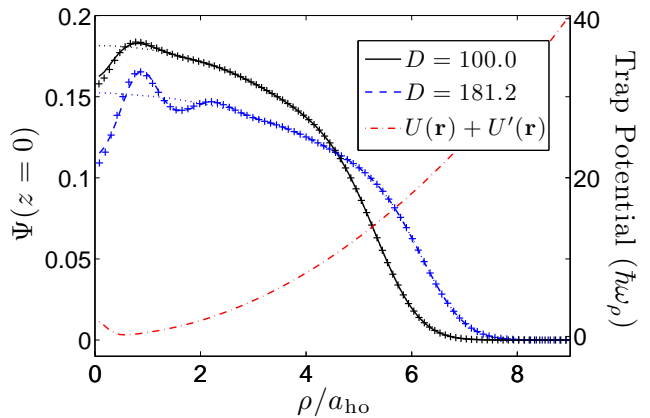}
\caption{\label{fig:profiles} Radial profiles of the $k=0$ dBEC
subject to the perturbing potential $U'(\rr) = \hbar \omega_\rho
\exp{(-\rho^2 / 2 (.2\, a_\mathrm{ho})^2)}$ in a trap with aspect ratio
$\lambda = 17$.  The red dash-dotted line represents the trapping
potential at $z=0$, the black solid line represents the radial profile
of the dBEC at $D=100$ and the blue dotted line represents the
radial profile at $D=181.2$, near the point of dynamic instability for
the $k=0$ dBEC.  The ``$+$'' signs represent the perturbation theory
results and the thin dotted lines represent the unperturbed radial
profiles at the corresponding dipole strengths.}
\end{figure}

For sufficiently oblate traps, $k=0$ dBECs exhibit radial density
oscillations in the presence of such Gaussian potentials.  Fig.~\ref{fig:profiles} illustrates the radial profiles of $k=0$ dBECs in a
harmonic trap with aspect ratio $\lambda = 17$ and with a Gaussian
potential having $A = \hbar \omega_\rho$ and $\rho_0 = 0.2\,
a_\mathrm{ho}$.  To give a concrete example, for $^{52}\mathrm{Cr}$
atoms in a harmonic trap with radial frequency $\omega_\rho =
2\pi\times 100\,\mathrm{Hz}$, this translates to having a beam width
of $\rho_0 = 280 \, \mathrm{nm}$.  In this trap, an interaction strength of
$D=181.2$, very near the point on instability for a $k=0$ dBEC in a
trap with the above aspect ratio, may be achieved with $\sim104,000$
$^{52}\mathrm{Cr}$ atoms.  It is seen that in this case even a small
Gaussian perturbation makes a dramatic change in the dBEC density
profile. The radial oscillations near $D_\mathrm{crit}$ are much more
pronounced than for a smaller dBEC with $D=100$.  This is
suggestive of the roton presence in this structure, since the roton is
expected to emerge with increasing dipolar interaction strength (i.e,
increasing density for a fixed dipole moment)
\cite{Santos03a,Dalfovo92,Ronen06a}.  As was shown in reference~\cite{Ronen07}, the roton mode undergoes a significant decrease in energy with increase in $D$ until it achieves zero energy at $D_\mathrm{crit}$, marking the point of dynamical instability for the $k=0$ condensate.  Beyond this $D_\mathrm{crit}$, the roton energy is purely imaginary.  Examining the nature of the
roton itself within BdG theory tightens up its relationship with the
observed structure discussed above.

For a $k=0$ condensate, the coupled BdG equations reduce to a single equation, given by
\begin{equation}
\label{BdG}
\tilde{G} \tilde{F} \ket{f} = \omega^2 \ket{f}.
\end{equation}
Here, $\tilde{G} = P(G-\mu)P$ and $\tilde{F} = P(F-\mu)P$, where $P = \mathrm{I}-\ket{\Psi}\bra{\Psi}$ is the projection operator into the space orthogonal to ground state wavefuncion $\ket{\Psi}$.  Also, $G = H_0 + C$ and $F = H_0 + C + 2\,X$, where $H_0$ is the
zero-interaction Hamiltonian, $C$ describes a direct interaction and
$X$ describes an exchange interaction between the Bogoliubov
quasiparticle with eigenvector $\ket{f}$ and the condensate.  All of these
operators are defined as in reference~\cite{Ronen06a}.  The
eigenvector $\ket{f}$ is given by $\ket{f} = \ket{u} + \ket{v}$, where $\{ u,v \}$ are the
familiar Bogoliubov eigenfunctions.  The $\omega$ appearing on the
right hand side of Eq.~(\ref{BdG}) is the energy eigenvalue
corresponding to $\ket{f}$.

In Eq.~(\ref{BdG}), it is understood that the linear space on which $\tilde{F}$ and $\tilde{G}$ act, and to which $\ket{f}$ belongs, is orthogonal to $\ket{\Psi}$.  Thus, we eliminate a non-physical solution with eigenvalue zero~\cite{Huepe03}.  The justification for working in this reduced linear space is that it can be shown that all physical excitations obey $\braket{f}{\Psi}=0$~\cite{BEC2003}.

It seems natural to assume that 
% like in ordinary time-independent perturbation theory%, 
the roton mode dominates the structure of the perturbed dBEC near instability because its energy
is much lower than the energies of the other BdG modes.  
%However, BdG theory by itself is not equipped with the formalism to make such statements. For this%  
To explicitly demonstrate this, one needs to formulate a perturbation
theory of the nonlinear GPE with respect to external potential perturbation.

To do so, we begin by writing a perturbation to the trapping potential
as $U \rightarrow U+U'$, where $U'$ is the small
perturbation.  The response of the condensate wavefunction to this perturbation is then $\ket{\Psi}
\rightarrow \ket{\Psi}+ \ket{\Psi'}$.  We insert these expressions into Eq.~(\ref{GPE}),
linearize in the primed quantities, and obtain the equation
\begin{equation}
\label{lambda}
\tilde{F} \ket{\Psi'} = -P U' \ket{\Psi}.
\end{equation}
To solve Eq.~(\ref{lambda}), we introduce a basis
defined by the eigenvalue equation
\begin{equation}
\label{eigenvalue}
\tilde{F} \ket{\varphi_n} = \varepsilon_n \ket{\varphi_n}
\end{equation}
and use its eigenfunction solutions to expand $\ket{\Psi'}$ in
the $\ket{\varphi_n}$ basis.  Plugging these expansions back into
Equation (\ref{lambda}) and working to first order gives
the expression for the wavefunction perturbation,
\begin{equation}
\label{psipert}
\ket{\Psi'} = -\sum_n  \frac{\bra{\varphi_n} U' \ket{\Psi}}{\varepsilon_n} \ket{\varphi_n}_.
\end{equation}
This derivation involves the use of the orthogonality condition
$\braket{\Psi'}{\Psi}=0$ and the fact that $\braket{\varphi_n}{\Psi}=0$.  The final expression is formally identical to that of the usual perturbation theory of the linear Schr\"odinger equation.

The connection between the BdG roton mode and the perturbative modes
is clear in the limit that the roton mode becomes degenerate with the
ground state.  In this limit, the roton energy $\omega$ goes to zero.
In Eq.~(\ref{BdG}), this means that $\tilde{G}\tilde{F}$ has
eigenvalue zero.  Now, note that the operator $G$ is positive semi-definite (its lowest eigenvalue is zero, with eigenfunction $\ket{\Psi}$.  This is indeed the ground state, since $\ket{\Psi}$ is nodeless).  Accordingly, the operator $\tilde{G}$ that, by definition, acts on the linear space orthogonal to $\ket{\Psi}$, is positive definite.  It then follows that any solution of $\tilde{G}\tilde{F}\ket{f_\mathrm{roton}} = 0$ must also satisfy $\tilde{F}\ket{f_\mathrm{roton}}=0$.  Thus, $\ket{\varphi_0}=\ket{f_\mathrm{roton}}$ is a solution of Eq.~(\ref{eigenvalue}) with eigenvalue $\varepsilon_0 = 0$.  Since $\ket{\Psi'}$ is written as an
expansion in $\ket{\varphi_n}$ with weights proportional to
$1/\varepsilon_n$, the eigenfunction $\ket{\varphi_0}$ with
eigenvalue $\varepsilon_0 \sim 0$ makes a contribution to $\ket{\Psi'}$
that strongly overwhelms the contributions of the other
eigenfunctions. Thus, in the limit that the roton energy goes to zero,
$\ket{\Psi'}$ is dominated by the BdG roton mode, $\ket{f_\mathrm{roton}}$.

\begin{figure}
\includegraphics[width=\columnwidth]{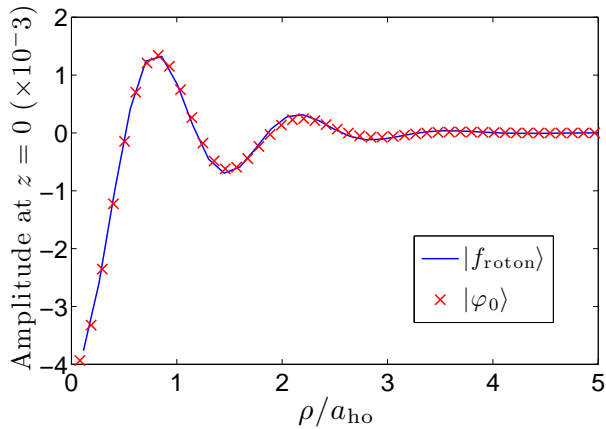}
\caption{\label{fig:ex} Radial profiles of excitations on a
rotationless dBEC with dipole strength $D=181.2$ in a trap with aspect
ratio $\lambda = 17$.  The solid blue line represents the BdG roton
mode while the red marks represent the $F$-operator eigenfunction with
eigenvalue $\mu$, $\ket{\varphi_0}$.}
\end{figure}

 To show that $\ket{\varphi_0}$ becomes identical to BdG
roton mode $\ket{f_\mathrm{roton}}$ when the roton energy goes to zero, Fig.~\ref{fig:ex} shows the radial profiles of both of these excited modes for a
rotationless dBEC with dipole strength $D=181.2$ in a trap with aspect
ratio $\lambda=17$, which is very near the point of instability.  Additionally, Fig.~\ref{fig:profiles} illustrates the accuracy with which this
perturbation theory predicts the wavefunction of a dBEC when perturbed
by a Gaussian potential, as discussed earlier in this Letter.
% In
%addition to the case of $\rho_0 = 0.2\,a_\mathrm{ho}$ shown in this
%figure, we have studied a variety of other Gaussian widths and have
%found that the perturbation theory is very accurate for widths of
%$\rho_0 \lesssim 0.7 \, a_\mathrm{ho}$, as long as their amplitudes
%remain small relative to the trapping potential.

 Recall that the $k=1$ solution of the GPE gave rise to a centrifugal
 potential in the radial part of Eq.~(\ref{GPE2}).  This
 potential 
%like the Gaussian potential produced by a blue-detuned laser%
 is constant along the trap axis and decreases quickly in the radial
 direction.  So, just as the Gaussian potential perturbs 
%the otherwise smooth % 
 the dBEC and gives rise to ripples on its density profile, we
 expect similar behavior for trapped dBECs with a centrifugal
 potential, i.e., dBECs with vortex structure.  To treat the
 centrifugal potential with our perturbation theory, we introduce a
 radial cutoff that is chosen to be much smaller than the spatial
 extent of the vortex core itself.  We find that for large $\lambda$
 there is good agreement between our perturbation theory and the
 results of our exact calculations.  Just as is the case for a
 Gaussian perturbing potential, the roton mode is responsible for the
 rich structure observed in the $k=1$ vortex state of a dBEC close to
 instability.
 
In conclusion, we have developed a perturbation theory for the GPE and
have applied it to dBECs perturbed both by thin gaussian potentials
centered on the trap axis and centrifugal potentials.  This theory
allows us to relate the radial oscillations observed on the exact
ground state profiles of perturbed dBECs to the roton mode observed in
the BdG spectrum of rotationless dBECs. For $^{52}\mathrm{Cr}$ and the
trap parameters discussed in this Letter, the length scale of the
oscillations is $\sim 2\mathrm{\mu m}$. This is in comparison to the
length scale of the predicted ripple in the $^4\mathrm{He}$ vortex, which is of
the order of 1 \AA, and has not been resolved experimentally up to
now.

The authors would like to acknowledge the financial support of the DOE and NSF.

\end{document}